\documentclass[11pt]{article}
\usepackage[latin1]{inputenc}
\usepackage{epsfig}
\usepackage{amsmath,amsfonts,latexsym, amssymb}

\newtheorem{satz}{Theorem}[section]
\newtheorem{defi}[satz]{Definition}

\newtheorem{koro}[satz]{Corollary}

\newtheorem{assumption}[satz]{Assumption}

\newtheorem{conclusion}[satz]{Conclusion}
\newtheorem{ob}[satz]{Observation}

\newtheorem{postulate}[satz]{Postulate}
\newtheorem{conjecture}[satz]{Conjecture}

\newcommand{\mcal}{\mathcal}
\newcommand{\mbf}{\mathbf}

\newcommand{\tit}{\textit}

\begin{document}
\thispagestyle{empty}
\begin{center}
\vspace*{1.0cm}

{\LARGE{\bf The Statistical Mechanics of Microscopic\\ Long-Range Bulk-Boundary
Dependence\\ in
 Black-Hole Physics and Holography}} 

\vskip 1.5cm

{\large {\bf Manfred Requardt }} 

\vskip 0.5 cm 

Institut f\"ur Theoretische Physik \\ 
Universit\"at G\"ottingen \\ 
Friedrich-Hund-Platz 1 \\ 
37077 G\"ottingen \quad Germany\\
(E-mail: requardt@theorie.physik.uni-goettingen.de)

\end{center}

\vspace{0.5 cm}

\begin{abstract}
  We argue in the following that the entropy-area law of black-hole
  physics and the various holographic bounds are the consequences of
  the microscopic dynamics of elementary degrees of freedom living on
  or near the Planck scale. We locate them both in the interior and on
  the boundary of, for example, the black hole with the strange
  area-behavior of various quantities being the result of a long-range
  bulk-boundary dependence among these degrees of freedom. In contrast
  to other approaches we regard the vacuum fluctuations on microscopic
  scales as the relevant elementary building blocks. In so far certain
  relations to to old ideas of Sakharov, Zeldovich et al are
  acknowledged (induced gravity). Most importantly, we prove that the
  existence of a large energy gap between a few low-lying excitation
  patterns and the majority of the other (in principle) possible
  excitation patterns in a subvolume with given boundary excitation is
  crucial for this area-dependence. We also remark that this is an
  indication that some particular entangled space-time geometry of a
  somewhat non-local character prevails in the microscopic (Planck)
  regime. Our findings are corroborated by the explanation of a number
  of open questions in the field (see the table of contents at the end
  of the introduction).
\end{abstract} \newpage
\setcounter{page}{1}
\section{Introduction}
In \cite{Bekenstein2} Bekenstein remarked that the deeper meaning of
black-hole entropy (henceforth abbreviated by BH) remains mysterious.
He asks, is it similar to that of ordinary entropy, i.e. the log of a
counting of internal BH-states, associated with a single BH-exterior?
(\cite{Bekenstein1},\cite{Bekenstein3} or \cite{Hawking1}). Or,
similarly, is it the log of the number of ways, in which the BH might
be formed. Or is it the log of the number of horizon quantum states?
(\cite{Hooft1},\cite{Susskind1}). Does it stand for information, lost
in the transcendence of the hallowed principle of unitary evolution?
(\cite{Hawking2},\cite{Giddings}). He then claims that the usefulness
of any proposed interpretation of BH-entropy depends on how well it
relates to the original ``statistical'' aspect of entropy as a measure
of disorder, missing information, multiplicity of microstates
compatible with a given macrostate, etc.

We think, the latter statement is a very important remark which, in
our view, is sometimes lost sight of in the discussion. It is in
particular the dynamical aspect of disorder which is important in
statistical mechanics and which goes beyond the frequently invoked but
physically somewhat empty pure information-entropy point of view
(subjective ignorance). See in this context the remarks on p.4545 of \cite{Frolov1}. 

A large group of researchers in the field view the question of the
localisation of the microscopic degrees of freedom (henceforth DoF),
generating BH-entropy in a seemingly rather geometric way and locate
them on or in the vicinity of the BH-horizon (to mention a few,
\cite{Hooft2},\cite{Sorkin1},\cite{Sorkin2}). In e.g. \cite{Sorkin1}
thesis 1 reads:``\ldots S resides on the horizon'', while in thesis 4 it
is stated: `` The idea is wrong that the DoF are inside the BH''. An
argument to this effect is given for example in \cite{Sorkin2} and
goes as follows:``\ldots The coupling from outside to inside is not weak
but very strong, while the reverse coupling is not so much weak as non
existent! Indeed this last observation points up the fact that
conditions in the interior should be irrelevant, almost by definition,
to what goes on outside.\ldots that it should have anything to do with
counting interior states''.

As to the latter point, we must admit that we are a little bit
sceptical as, in our view, it seems to be a classical or
quasiclassical argument. We will show that at or near the
Planck-scale, which is in our approach (see below) the appropriate
environment, there exists a marked correlation connecting the interior
and the exterior of the BH through the horizon. We already provided
strong arguments in favor of the existence of such a non-local and
long-range collective behavior in the second part of \cite{Requ1} and
enter into a quite detailed quantitative analysis of the microscopic
correlation structure in section \ref{Influence} of the present paper.
We argue that making a few reasonable assumptions about the fine
structure of the vacuum fluctuation spectrum which can be
observationally confirmed and amalgamating this with the holographic
principle in scenarios where the latter can be confirmed, we can
rigorously prove that the vacum fluctuations are long-range
(anti)correlated in (quantum)space-time on microscopic scales (see the
following for more details).

The idea that the relevant DoF can be thought of as being essentially
confined to a thin halo about the horizon, may have been inspired by
the apparent behavior of the physical concept of
\tit{entanglement-entropy}.  This is a concept with wide applications
also in statistical mechanics and/or quantum information theory (see
e.g. \cite{Requ2} for more references); the two papers which are
relevant in our present context are \cite{Sorkin3} and
\cite{Srednicki}; see also the more recent \cite{Plenio}. In these
papers certain arrays of coupled harmonic oscillators and their
continuum limit (a Klein-Gordon field theory) have been analysed.
After some delicate and tedious calculations it was shown that the
entanglement-entropy of the groundstate (i.e. the vacuum in the
continuum limit), when traced over a subvolume, $V_i$, with $V=V_1\cup
V_2$, is proportional to the area of the dividing surface while the
correct calculation of the prefactor is more delicate. These
calculations were however performed in flat Minkowski space but it was
argued that the results have a certain bearing also for the
BH-situation. As a consequence of these findings there now seems to be
a certain tendency to associate BH-entropy (at least to some
degree) with entanglement-entropy and having its origin in the DoF
near the horizon.

To put these various results in perspective we would like to make the
following points clear. In \cite{Requ2} we showed that under the
condition that interactions and correlations are short-ranged,
groundstates lead to an entanglement-entropy which is proportional to
the area of the dividing surface for a wide range of Hamiltonians. On
the other hand, for systems being in a state which displays long-range
correlations (e.g. a \tit{(quantum) critical state}) this does not
hold even for the groundstate. This shows that the assumption of
short-range correlations is important. Furthermore, even in the
short-range case, for eigenstates which are highly or lowly excited,
we proved that the entropy depends linearly on the volume or the log
of the volume of the subsystems times the area of the dividing
surface. This implies that in these latter situations the entropy is
no longer localized near the dividing boundary.

This is supported by yet another interesting result. With the state of
a subsystem, defined over a subvolume, $V_1$, being a canonical Gibbs
state, we can extend this temperature state to a pure vector state,
$\Psi$, on the larger volume, $V_1\cup V_2$, in such a way that its
restriction, i.e. its \tit{partial trace}, to $V_1$ is the Gibbs
state, we started from. This implies that the
\tit{entanglement-entropy} of the total vector state, $\Psi$, with
respect to $V_1$ is the original thermodynamic entropy of the Gibbs
state. The latter happens to be proportional to the volume $V_1$ in
the generic case, which by the same token holds for the
entanglement-entropy of $\Psi$. We thus see that quite a lot of vector
states have entanglement-entropies being proportional to the
respective subvolumes. In this context see also the observations in
\cite{Wehrl} section II.D and \cite{Plenio2}

What does this mean for the understanding of BH-entropy? Due to the
work of Bekenstein and many others we have learned that the maximal!
entropy or information which can be stored in a BH is proportional to
the area of the horizon (which is a stronger result as the one
referring only to some particular state). Making an educated
speculation, we model the interior of the BH as some kind of
(quantum)statistical subsystem on a sufficiently microscopic scale
with the local vacuum fluctuations on this presumed fundamental scale
as its degrees of freedom. This idea is actually not so far-fetched;
cf. the old ideas of e.g. Sakharov and Wheeler (see below).  It should
then be possible in principle to put this system in a higher excited
state by making e.g. the fluctuation spectrum less correlated, i.e.
increasing its dissorder.  In case the fluctuation spectrum happens to
be short-range correlated, our above cited rigorous results
(\cite{Requ2}) show that the maximal entropy should be proportional to
the volume and not! the area. We hence arrive at the preliminary
result
\begin{conclusion}The assumption of short-range correlations is
  necessarily violated in the BH-context. On the other hand, the
  conclusion that entanglement-entropy is proportional to the area was
  (among other things) based on this (tacit) assumption. But even in
  the case where this situation would prevail, the result does not
  hold for the maximal! entropy.  Therefore we conjecture that an
  explanation of BH-entropy is more delicate and seems to need some
  more prerequisites.\end{conclusion}
 In the following we want to provide some of these missing
 prerequisites and undertake to formulate a theory of
 \tit{bulk-boundary-statistical mechanics} which extends to some
 extent ordinary statistical mechanics.

 There is a thoughtful discussion of such problems in \cite{Wald1}
 p.31ff which points in a similar direction concerning the problem of
 the localisation of the responsible DoF. It is argued that, in the
 end, all the different suggestions like e.g. thermal atmosphere,
 horizon, interior, may come out to be complementary aspects of the
 same physical DoF, a working hypothesis we try to substantiate in the
 follwing. Two other well-written papers of partly review character
 and dealing with these questions and the respective context are
 \cite{Jacobson1},\cite{Bousso}. In \cite{Bousso}, for example, the
 idea is contemplated that pure information may underlie ultimately
 all of physics. This point of view is corroborated by the observation
 that the holographic principle betrays little about the character of
 the microscopic DoF (\cite{Bousso}). We are following a similar
 working philosophy for already quite some time (see e.g.
 \cite{Requ1},\cite{Requ3},\cite{Requ-Gromov},\cite{Requ-Wormhole},
 \cite{CA} and further earlier references given there).

 This whole approach relates in an interesting way to other
 fundamental theories being in vogue presently as e.g. string theory
 or loop quantum gravity, which are not really of this mentioned
 character. There may be two modes of relation between these, at first
 glance, not entirely compatible approaches. Either, the theory we are
 developing in the following, and which builds on the excitation
 patterns of vacuum fluctuations on primordial scales as fundamental
 building blocks (extending old ideas of Sakharov et al), lives on a
 finer scale than these two other theories, which have then to be
 regarded as derived theories. Or, on the contrary, this framework is
 itself kind of an effective theory which uses some gross features of
 a more fundamental theory. Both cases are in principle possible while
 we personally favor the first alternative. Anyhow, the phenomenon
 that the entropy-area law can apparently be derived both in string
 theory and loop quantum gravity for at least certain extremal cases,
 and, on the other hand, can also be understood in our more model
 independent framework, may be an indication that it is perhaps the
 consequence of somewhat more general features, a point of view we
 explicitly forward in our approach.

 As the paper is relatively long and deals with quite a variety of
 subjects which are grouped around some central sections, we conclude
 this introduction, for convenience of the reader, with a table of
 contents. A central role is played by the quantitative analysis in
 section \ref{4}, while in the sections after section \ref{4}, the
 preceding results are applied to a variety of important questions in
 the field.
\tableofcontents
\section{Some Fundamental Issues}
One reason why the question of the microscopic origin of BH-entropy is
still not yet settled after so many years of intense discussions lies
in the fact that this topic is (so to speak) situated just at the
horizon between a region in ``theory space'' which is understood at
least in principle, i.e. quantum field theory (Q.F.Th.) in curved
space-time, and a region which is still sort of a terra incognita,
i.e. quantum gravity (Q.Gr.) (irrespective of the claims of string
theory). Therefore most attempts tried to attack the open problems
from the better known exterior region (the low-energy end) by
exploiting the methods and tools of Q.F.Th. and, consequently, located
the microscopic origin of BH-entropy mainly in the region exterior to
the horizon or the horizon itself (either thermal atmosphere or
horizon states or both).

One feature which appears in traditional Q.F.Th. is the existence of
zero-point oscillations (or excitations) of the Fourier modes in the
expansion of, for example, the Hamiltonian of free quantum field
theories. These excitations were frequently neglected to some extent
(``renormalized away'')and viewed as ``unphysical'' but have come to a
certain prominence again in the context of BH-entropy and related
fields like e.g. the Casimir or the Unruh effect where they prove
their objective existence. We think that the (re)emergence of these
notions is pointing in the right direction but are sceptical if they
can really provide a quantitatively correct explanation if dealt with
in the usual, somewhat narrow traditional way. First, these virtual
vacuum excitations typically show up in a perturbational treatment of
Q.F.Th.  in a specific and model dependent way. This implies that a
Klein-Gordon theory vacuum looks, so to speak, different from a
QED-vacuum etc., each having its particular vacuum excitation modes.
In principle, the real vacuum outside a BH should then support all
these different excitation modes and in addition all the vacuum
excitations showing up in all the effective (and at the moment mostly
unknown) theories living on higher energy scales up to the Planck
scale and, last but not least, all the geometric and possibly
topological excitations of space-time itself.  After all, this seems
to represent a great higgeldy-piggeldy of superficially different
excitation patterns, while we think, given the well-known limited
value of the particle picture in fully developed Q.Gr., that they will
look all alike near the Planck scale and supposedly do no longer have
a (virtual) particle character of whatever type.

The history of the idea of vacuum fluctuations and/or zero-point
energies is both involved and fascinating. Nice recent historical
reviews are \cite{Zinkernagel1} and \cite{Zinkernagel2}. It is
particularly noteworthy that already Nernst discussed in quite some
detail the concept of zero-point energy which he considered as being
situated in the aether (\cite{Nernst}). He even had the idea that
energy conservation may only hold in a statistical sense and that
particles do perform what was later called \tit{Zitterbewegung}.

Pauli came back to the problem and related it (presumably for the
first time) to general relativity and the \tit{cosmological constant}
(but apparently did not realize the necessity of a negative pressure!,
which has a repulsive effect, as he stated that his calculations show
that the universe would not even stretch to the moon). For these
reasons he seemed to ignore the possibility of zero-point energies
(cf. the remarks in \cite{Enz}). By and large, zero-point energies
were not taken very seriously at that time anyhow, as can be inferred
from a letter by Bohr to Pauli (quoted in \cite{Zinkernagel2}). In
that letter Bohr rightly remarks that in quantum theory such effects
can only be observed by making measurements, implying the interference
of quantum objects with \tit{macroscopic} objects. From a strict
logical point of view one can therefore never decide if e.g. field
fluctuations were already present in the pure vacuum or, on the other
hand, have been created by the interference with a measuring
apparatus. We should emphasize that this ambiguity besets the whole
field of quantum theory and is the reason why such fundamental
questions are difficult to settle once for all. It is here not the
place to enter into a necessarily involved epistemological debate. We
should however point to related epistemological problems in general
relativity. There the question is, is space-time ``really'' curved in
an objective sense or are, on the other hand, the measuring
instruments deformed by gravity. It is actually the (aesthetic)
question which theory is more \tit{coherent} and satisfying as
Einstein used to point out.

More recent reviews of the \tit{cosmological constant problem} and its
relation to vacuum fluctuations are e.g.  \cite{Boyer},\cite{Weinberg}
and \cite{Straumann}.  Concerning our own line of thought, a very
clearly written contribution is \cite{Zeldovich}. Zeldovich in
particular points out that one of the typical arguments that
zero-point energies are artefacts is wrong.  That means, the
(standard) reasoning that a quantum-field-system vacuum four-vector
with, say, energy $(E,0)$ must go over into some $(E',\mbf{p'})$ under
a Lorentz transformation with $\mbf{p'}\neq 0$ unless $E=0$ (i.e. it
is not Lorentz-invariant). This conclusion is not correct for various
reasons (actually, $E$ is infinite in our case ). But we should in
particular emphasize that in a complete theory including gravitation
energy-momentum is rather part of the energy-momentum two-tensor and
an object like $const\cdot Diag(1,-1,-1,-1)$ or rather $const\cdot
g_{\mu\nu}$ is covariant and yields the negative contribution to the
pressure. We can however conclude the following:
\begin{conclusion} We can learn from the preceding remarks that the
  inclusion of gravity is crucial if one wants to deal consistently
  with zero-point energies.   \end{conclusion}

To conclude this brief interlude about zero-point energies one should
make it clear that all the zero-point energies occurring in models do
arise from pure fluctuations of some observables, thus underpinning
(at least in our view) their real existence. In the most simple
example, the harmonic oscillator, the Hamiltonian is essentially the
sum of $P^2$ and $Q^2$, i.e.
\begin{equation}H=P^2/2m+m\omega/2\cdot Q^2     \end{equation}
and with 
\begin{equation} \langle P\rangle_0=\langle Q\rangle_0 =0
\end{equation}
in the groundstate, $\psi_0$, we have
\begin{equation}\hbar\cdot\omega/2=\langle H\rangle_0=1/2m\cdot
  \langle (P-\langle P\rangle_0)^2\rangle_0+ m\omega/2\cdot\langle (Q-\langle Q\rangle_0)^2\rangle_0      \end{equation}
with 
\begin{equation}\langle (P-\langle P\rangle_0)^2\rangle_0\cdot \langle
  (Q-\langle Q\rangle_0)^2\rangle_0\geq \hbar^2/4   \end{equation}
which follows from $[P,Q]=-i\hbar$.

In the same way we have in (matter-free) QED:
\begin{equation}H=const\cdot (\mbf{E}^2+\mbf{B}^2)   \end{equation}
with
\begin{equation} \langle \mbf{E}\rangle_0=\langle \mbf{B}\rangle_0 =0  \end{equation}
so that again $\langle H\rangle_0$ is a sum over pure vacuum
fluctuations of the non-commuting quantities $\mbf{E}$ and $\mbf{B}$.

Another point is the problem of continuous space in general in
connection with, for example, entanglement-entropy. It is easy to make
relatively rigorous calculations within the framework of quantum
lattice theories (see e.g. \cite{Requ2}). It is in particular
unproblematical to divide a quantum system into two subsystems in this
framework. The total Hilbert-space becomes the natural tensor product
of the two subspaces while the total algebra of observables can be
uniquely represented as the tensor product of the corresponding
subalgebras., i.e.
\begin{equation}\mcal{H}=\mcal{H}_1\otimes \mcal{H}_2\qquad
  \mcal{A}=\mcal{A}_1\otimes \mcal{A}_2    \end{equation}  
This is certainly the reason why both \cite{Sorkin3} and
\cite{Srednicki} start from a discretized version of a continuous
theory. Taking then the continuum limit is difficult and not free
of ambiguities since infinities do arise which are not so easy to get
rid of in a non-adhoc manner.

On the other hand, when starting directly from a continuous Q.F.Th.
model, it is not obvious in the general situation how to make a
natural division of this model system into two subsystems (without
losing certain contributions!), e.g. divide a total Fock space into a
tensor product of the Fock spaces of two subsystems. For special
situations this can be done in a reasonable way like the Rindler wedge
(see the seminal paper \cite{Fulling1} or \cite{Fulling2} for a
concise discussion), but in general appropriate explicit coordinate
transformations are not at our disposal which allow to represent a
subsystem with the help of two different mode expansions, the one
belonging to the total system, the other being restricted to the
subspace. In this context the existence of \tit{time-like
  Killing-vector fields} play an important role.\\[0.3cm]
Remark: We mention in passing the important results of axiomatic
Q.F.Th. which show how difficult it is to isolate subsystems, defined
by local algebras of observables and the physical consequences of
Tomita-Takesaki-theory (\cite{Haag}).\vspace{0.3cm}

Another point which is not really clear to us is the range of validity
of this Q.F.Th. approach when it comes to the really microscopic DoF
of the quantum vacuum, i.e. when we approach the Planck-scale, as we
surmise that the bulk of DoF contributing to e.g. the BH-entropy is of
this very microscopic nature. Furthermore, in the continuum approach
one has to introduce an adhoc UV-cutoff in order to keep the
calculated value of the entropy finite. Note in this respect the
choice of a ``brick wall'' in \cite{Hooft2} in order to make an
effective division of interior and exterior of the BH. It would be
advantageous in our view to have a framework which generates such a
cutoff in a natural way.

In the following we undertake to develop such a ``bottom-up''
approach. We start from a microscopic regime in which higher complex
structures like e.g. quantum fields and particle excitations emerge as
derived and extended excitation patterns living on this primordial
array of elementary DoF (see also \cite{Requ1} and further references
given there). In this respect we want to mention the old ideas of
Sakharov (\cite{Sakharov1},\cite{Ruzmaikin},\cite{Sakharov2} or
\cite{Wheeler1}, p.426ff). To put it briefly, Sakharov argued that for
example the gravitational field is a derived effect of deformations in
the spectrum of vacuum fluctuations, thus yielding sort of a ``metric
elasticity of space''; in \cite{Jacobson2} this is called \tit{induced
  gravity}, see also \cite{Adler}. This represents in our view a deep
shift in emphasis concerning the fundamental constituents of our
understanding of the structure of our hierarchy of physical theories.
I.e., the traditional order of analysis is reversed. Instead of
starting from particles or fields and their interaction and frequently
only indirectly and at a later stage, arriving at the analysis of e.g.
the fine structure of the physical vacuum, we do not regard the vacuum
as sort of a stage but rather as the real source of all the objects
and phenomena emerging in it.

We developed and described such an approach in the papers, mentioned
in the introduction. As a further motivation we would like to cite a
passage from \cite{Wheeler1}, p.2202f, just to show that such a point
of view is by no means entirely far-fetched:``\ldots A particle means
as little to the physics of the vacuum as a cloud means to the physics
of the sky. In other words, elementary particles do not form a really
basic starting point for the description of nature. Instead they
represent a first-order correction to vacuum physics. The vacuum, that
zero-order state of affairs, with its enormous densities of virtual
photons and virtual positive-negative pairs and virtual wormholes, has
to be described properly before one has a fundamental starting point
for a proper perturbation-theoretic analysis''.

As to our concrete enterprise, we adopt the following strategy. We
assume certain qualitative and, as we think, well-founded properties
concerning the underlying microscopic substratum and which are
expected to hold in any case, irrespective of the concrete shape of a
possible future theory of Q.Gr. In so far the approach is the same as
in \cite{Requ1} and in some of the papers cited there. In a sense one
may describe this working philosophy as being similar to a foundation
of phenomenological thermodynamics which is based on certain general
properties of a supposed underlying theory of statistical mechanics
but not on any details or models of the latter (as it is e.g. done in
\cite{Callen}). One may it also call a \tit{principle theory} in the
sense of Einstein (cf. \cite{Einstein}), that is, one postulates
certain general principles, to start from, without making too detailed
or uncontrolled model assumptions.

To be more specific, by making a few and observationally well-founded
assumptions about the behavior of the spectrum and correlations of
vacuum fluctuations, we derive a couple of almost model-independent
and general results which lead to surprisingly strong constraints for
the physics on this microscopic scale. In this way we hope to infer
general and structural results about a regime to which we, at the
moment, do not have direct experimental access. By the way, we think
it is perhaps a funny side remark that in both fields, i.e.
phenomenological thermodynamics and our BH-context the almost
universal concept of entropy is the really crucial analytic tool to
infer some deep results about an underlying microscopic theory, the
details of which are not yet known or remain unresolved.
\section{The Fundamental Postulates}
We presume the two main reasons why the possibility that the
entropy-area law may be generated by DoF sitting inside the BH,
i.e. within the horizon, has never been seriously taken into account
in most of the representations, are the following:
\begin{itemize}
\item Superficially, the interior of a BH seems to be an essentially empty
and inactive area.
\item The area-law has been mainly viewed as an indication of
its geometric origin; entropy proportional to horizon area meaning:
DoF located on or near the horizon.
\end{itemize}

On the other hand, people with a stronger statistical-mechanical
background may come to a different conclusion. Given that in a first
approximation a large part of the BH interior (with the possible
exception of its central singularity and its immediate neighborhood)
is not entirely different from an arbitrary piece of empty space or
space-time being exposed to some gravitational field, we assume that,
apart from possible finite distortions (see e.g. the Unruh-effect),
the fluctuation spectrum of the vacuum excitations is at least
qualitatively similar in both cases.

In \cite{Requ1} we performed the following thought experiment. The
assumption that on very small scales the extent of vacuum fluctuations
is very large and extremely large if we approach the Planck scale,
even if perhaps not shared by every worker in the field, seems to be
corroborated by a wide spectrum of more or less independent
inferences. The following model assumptions seems in our view
therefore reasonable.   
\begin{postulate}It is allowed to replace a piece of space by a
  coarse-grained statistical model which is composed of microscopic
  grains of, to supply a typical scale, Planck-size which support
  elementary DoF which, individually, can strongly fluctuate. In
  energy units the elementary fluctuations are assumed to be of
  Planck-energy size.
\end{postulate}

In sect.4 of \cite{Requ1} we made a calculation which shows that,
given the huge number of roughly Planck-size grains in such a piece of
space and provided that the individual grains are allowed to fluctuate
almost independently, the total fluctuations in a macroscopic or
mesoscopic piece of space of typical physical quantities are still
large enough (i.e. macroscopic) as to be observable. More precisely,
with $q_i$ some physical quantity belonging to a microscopic grain
(e.g.  energy, momentum, some charge etc.) and $Q_V:=\sum_i q_i$ the
observable belonging to the volume $V$, the fluctuation of the latter
behaves as
\begin{equation}\langle Q_VQ_V\rangle^{1/2}\sim V^{1/2}\end{equation}
with $N\sim V$ the number of grains in $V$. This is a consequence of
the \tit{central limit theorem}. As such large integrated fluctuations
in a macroscopic region of the physical vacuum are not observed (they
are in fact rather microscopic on macroscopic scales), we conclude:
\begin{conclusion}The individual grains or supposed elementary DoF do
  not fluctuate approximately independently.
\end{conclusion}
Remark: We note that this fact is also corroborated by other,
independent observations.\vspace{0.3cm}

We can refine the result further (cf. \cite{Requ1}) by assuming that
the fluctuations in the individual grains are in fact correlated over
a certain distance or are \tit{short-range correlated}. In
mathematical form this is expressed as \tit{integrable correlations}.
This allows that ``positive'' and ``negative'' deviations from the
mean value can compensate each other. Letting e.g. $q(x)$ be the
density of a certain physical observable and $Q_V:=\int_V
q(x)\,d^n\!x$ the integral over $V$. In order that
\begin{equation}\langle Q_VQ_V\rangle^{1/2}\ll V^{1/2}
\end{equation}
we proved in \cite{Requ1} that it is necessary that
\begin{equation}\int_V \!d^n\!y\,\langle q(x)q(y)\rangle\approx 0 \label{fluc}   \end{equation}
Remark: For convenience we normalize $q(x)$ so that $\langle
q(x)\rangle=0$ with $\langle\circ\rangle$ denoting the expectation in the
vacuum state (or, in another context, a thermodynamic equilibrium
state).\vspace{0.3cm}

We made a more detailed analysis in \cite{Requ1} under what conditions
property (\ref{fluc}) can be achieved. In any case we can again conclude:
\begin{conclusion}Nearly vanishing fluctuations in a macroscopic
  volume, $V$, together with short-range correlations imply that the
  fluctuations in the individual grains are anticorrelated in a
  fine-tuned way, i.e. positive and negative fluctuations strongly
  compensate each other which technically is expressed by property
  (\ref{fluc}).
\end{conclusion}

We now come to implications derived from the so-called
\tit{holographic principle}. For the time being, we only deal with
situations where the \tit{spacelike} holographic principle holds. That
means for example, (quasi)static backgrounds. The reason is that, as
our approach develops a relatively new point of view concerning this
context, we would like to keep the scenario in a first step free from
additional technical complications. We will howeveer briefly comment
upon dynamical aspects and time-dependent backgrounds in subsection
\ref{range} where we discuss a variety of examples. The range of its
validity is discussed in e.g. \cite{Bousso}; see also \cite{Fischler}
and \cite{Easther}. We note however that our following analysis will
shed a new light on this principle and its true range of validity and
will, furthermore, unearth presumably interesting relations to the
ideas of Zeldovich and Sakharov mentioned above.

 The BH-scenario in asymptotic flat space-time belongs to this class.
This leads to the next postulate.
\begin{postulate}There exists a class of scenarios in which the
  maximal amount of information or entropy which can be stored in a
  spherical volume is proportional to the area of the bounding
  surface. This is the spacelike holographic principle.
\end{postulate}
Remark: In some treatments this property is translated into the
statement that the number of available DoF is of order (area) or, even
stronger, the relevant DoF are situated on the bounding surface. In
our view this conclusion is wrong or at least premature as we try to
show in the following.\vspace{0.3cm}

Putting now together the content of postulate 3.1 and postulate 3.4,
we already reasoned in \cite{Requ1} that what all this is really
implying is that the DoF in the volume, $V$, are long-range
(anti)correlated in a very peculiar way (cf. also certain remarks in
section 7 of \cite{Susskind2}).
   \begin{conclusion}From postulate 3.4 we infer that
  each fluctuation pattern in $V$ is fixed by the corresponding
  pattern situated on the bounding surface or in a thin shell about
  this surface.
\end{conclusion}
This now leads to our final conclusion:
\begin{conclusion}The fluctuation pattern in $V$ is long-range
  anticorrelated in a fine-tuned way on a microscopic scale and is
  essentially fixed by the state of the fluctuations on the bounding
  surface.
\end{conclusion}
Remark: The peculiar relation between long-range-anticorrelations on a
microscopic scale and the correlations among extended excitation
patterns on a larger (e.g. mesoscopic) scale like e.g. ordinary
quantum mechanics is discussed in section \ref{mesoscopic}\vspace{0.3cm}

What remains to be done in a next step is to clarify the subtle
details of this fluctuation structure. This is a non-trivial task as
it turns out that, while the phenomenon of long-range anticorrelations
as such is certainly an important property, it is not! \tit{the}
crucial and characterizing property in this specific context. What is
really peculiar is the fixation of the bulk DoF by the surface DoF as
this latter property is not already implied by the former correlation
result. We will show that to achieve this we need yet another
prerequisite.

But before we will do this we want to show that the preceding
reasoning already allows us to draw some simple but important
geometric conclusions. For convenience we assume that the sets of DoF,
occurring in the following are countable. We now consider two
concentric spheres, $S_1,S_2$ with radii $R_1< R_2$. The DoF on $S_1$
by assumption determine the DoF inside $S_1$. The DoF on $S_2$
determine the DoF inside $S_2$ and in particular on $S_1$. Put
differently, in for example three dimensions, we have
\begin{equation}\#(\text{modes in} B_i)=const\cdot R_i^2
\end{equation}
where $B_i$ is the interior of $S_i$.

Furthermore there are $(R_2/R_1)^2$ different modes on $S_2$ per mode
on $S_1$. We denote the individual modes on $S_i$ or in $B_i$ by
$\omega_j^{S_i}$ and $\omega_j^{B_i}$ respectively, with mode
$\omega_j^{B_i}$ uniquely fixed by mode $\omega_j^{S_i}$. It follows
that to each $\omega_j^{S_1}$ corresponds a class,
$[\omega_j^{S_1}]^{S_2}$, of modes on $S_2$ with cardinality
$(R_2/R_1)^2$. By the same token, these latter modes fix
$(R_2/R_1)^2$ different modes in $B_2$. As the modes in $B_1$ are
standing in a 1-1-relation to modes on $S_1$, it follows 
\begin{ob}The modes in $B_2$, induced by the class
  $[\omega_j^{S_1}]^{S_2}$ are different from each other in
  $B_2\backslash B_1$ but coincide within $B_1$.
\end{ob}

\section{\label{4}The Statistical Mechanics of Bulk-Boundary Dependence}
Superficially considered one may be inclined to think that the
property of long-range anticorrelation is already sufficient to
understand the entropy-area-law. An instructive example (standing
however for many others), taken from the statistical mechanics of
phase transitions, shows that this is not the case (see subsection
\ref{4.2}) . But what is perhaps more important, we can learn from
this typical example what additional property is needed to yield this
strong and quite unusual result when looked upon with a statistical
mechanical eye (subsection \ref{4.3}). But before we come to that
issue, we want in a first step to clarify in more quantitative detail
the type of long-range correlation or influence and in particular its
weak spatial decay which seems to be prevalent in this context. We
emphasize that we take some pains to approach this question without
making adhoc assumptions like e.g. assuming some particular underlying
dynamical model. We only will employ the facts we have described
above. The results will have, among other things, some bearing on the
range of validity of the various holographic bounds discussed in
section \ref{appli}.
\subsection{\label{Influence}The Long-Range Character of Influence and
  Correlations on the Primordial Scale}
In this subsection we analyse the fine structure of the vacuum
fluctuations on the Planck scale and their correlations under the
condition that the spatial holographic principle holds in certain
situations. The quantitative results will however also explain where
and why the simple spatial holographic bound does not hold.  The
possible physical reasons for this exceptional behavior are then
isolated in the other subsections.

In principle there are two possible modes of discussion, first: in a
more classical statistical way, second: relying more on the arsenal of
tools provided by quantum (field) theory. In the litterature one
frequently finds a mixture of arguments taken from both fields, even
in the same paper. This is however not entirely unreasonable. For one,
we are mainly talking about model theories which are designed to
elucidate some typical behavior, not about a final fundamental theory.
For another, it is by no means clear that quantum theory still holds
sway unaltered on those fundamental scales we are interested in. After
all, it may only be yet another \tit{effective theory} which is only
correct within certain boundaries as to resolution of space-time.

This being said, we will analyse the above question mainly with the
help of ordinary probability theory. We think, the quantum case
behaves in a similar way but it is easier (for the time being) to
treat the DoF under discussion as ordinary \tit{random variables}
since they can be taken to commute among each other. Furthermore, it
is sometimes overlooked that a large part of ordinary quantum theory
can be cast into a form of ordinary probabilistic correlation analysis
(without for example employing Hilbert space methods), as has so
successfully been demonstrated by Bell (\cite{Bell}). 

That is, in the following we treat the DoF in a subvolume, $V$, plus
the ones being situated on the boundary, $\partial V$, as a
statistical system, the statistical behavior being mainly induced by
the openness of the systems relative to the outside regime (similar to
a \tit{heat-bath}). We restrict our analysis to a coarse-grained view
and treat the set of DoF as countably discrete with their respective
values also assumed discrete (but this is only a matter of
convenience).
\begin{defi}We regard the DoF as random variables living on a
  probability (or sample) space $\Omega$, the sample points, $\omega$,
  being individual fluctuation patterns at a given time in $V$ and/or
  $\partial V$ while the DoF are the elementary states of fluctuations
  located in certain grains, $\mcal{C}_i$ of this extended fluctuation
  pattern. The DoF in the interior, $V$, associated with the
  $\mcal{C}_i$ are denoted by $X_i$, the ones on the boundary by
  $Y_j$. Events, usually denoted by $A,\ldots$ are certain aggregates
  of sample points (excitation patterns), i.e. admissible (as to
  technical details of probability theory see below) subsets in
  $\Omega$.
\end{defi}

In the following, particular types of such sets are important. In
general probability theory or in the path integral formalism such sets
are called \tit{cylinder sets}. Let $\mcal{M}_{\nu}$ be an arbitrary
subset of elementary random variables $X_i$, i.e. DoF, and let their
respective momentary values be $x_i$,
\begin{defi}We denote by $A(\{\mcal{M}_{\nu}\ni X_i=x_i\})$ the subset
  of sample points $\omega$, i.e. fluctuation patterns with the
  local fluctuations (elementary DoF), $X_i$, in the grains $\mcal{C}_i$
  belonging to the selected set $\mcal{M}_{\nu}$ having the (discrete)
  values $\{x_i\}$.  We assume that the probability of such sets is
  non-vanishing provided the sets are not empty (see below), that is
\begin{equation}pr(A(\{\mcal{M}_{\nu}\ni
  X_i=x_i\}))>0\quad\text{if}\quad A\neq\emptyset   \end{equation}
\end{defi}
\begin{koro}For a single elementary random variable, $X$ (DoF), one
  can then define the associated probability distribution by
\begin{equation}d(x):=pr(X=x)   \end{equation}
with $x$ varying. By the same token we get its expectation value and
mean square deviation
\begin{equation}\overline{x}:=\sum_x x\cdot d(x)\quad ,\quad \Delta
  x^2:=\sum_x (x-\overline{x})^2\cdot d(x)  \end{equation}
\end{koro}
That is, the respective sets consist of excitation patterns in $V$
with the local fluctuation, $X$, in some $\mcal{C}$ having the fixed
value $x$. In the same vein one can define more general types of
events respectively subsets. For these and other elementary
probabilistic concepts see e.g. \cite{Renyi} or \cite{Feller}.

The following observation is important. The DoF on the boundary,
$\partial V$, already generate the full algebra of events in the interior.
\begin{ob}The holographic principle in the way we are using it in this
  paper tells us that a full specification of the momentary values of
  the DoF on the boundary, $Y_j=y_j$ for all $Y_j$ on $\partial V$ fix
  the values of all the $X_i$ in $V$. For our stochastic analysis this
  implies the following.\\
i) Each random variable, $X_i$, is a function of the set of $Y_j$,
i.e.
\begin{equation}X_i=F_i(\{Y_j\}_{\partial V})\quad , \quad
  x_i=f_i(\{y_j\}_{\partial V})   \end{equation}
and the $f_i$ define a map from the sample space $\Omega_{\partial V}$
to $\Omega_V$
\begin{equation}f:\{y_j\}_{\partial V}\;\to\; \{x_i=f_i(\{y_j\}\}
\end{equation} ii) By the same token, every event $A(\{X_i=x_i\}_V)$,
$X_i$ a single DoF, is either the one-point set
$A(\{Y_j=y_j\}_{\partial V})$ for some configuration $\{y_j\}$ on the
boundary if $\{x_i\}$ is lying in the image set of $f$ or it is empty.
In this way each general event in $V$ can be formed from corresponding
sets in $\partial V$ by using the usual set-theoretic constructions
employed in probability theory.
\end{ob}

We now want to estimate the strength of the statistical influence the
boundary DoF's are having on the DoF's in the interior. This turns out
to be a really subtle point. As we want to avoid adhoc model
assumptions, concepts like forces or direct interactions among the
DoF's are presumably too primitive, while the information encoded in
correlation functions is too limited for our purposes. Therefore, we
will introduce a new concept, which amalgamates the virtues of the
preceding concepts while avoiding the respective drawbacks. We will
call it \tit{influence function}. 

Remember that in previous sections we found that the DoF are
\tit{long-range correlated}. Traditionally, in statistical physics one
expresses such a property by means of \tit{correlation functions}. We
found however, that in our context, where we have only a
very particular (and limited) type of information at our disposal,
correlation functions are a too crude and inflexible tool as a
starting point (this will become clearer in the following analysis).
A, in our view, better starting point are \tit{conditional
  probabilities}. An elementary example is the following.
\begin{defi}We denote by $pr(X_i=x_i|Y_j=y_j)$ the probability of
  the event $X_i=x_i$ given $Y_j=y_j$. In more detail it reads
\begin{equation}pr(X_i=x_i|Y_j=y_j):=pr(X_i=x_i,Y_j=y_j)/pr(Y_j=y_j)     \end{equation}
provided that $pr(Y_j=y_j)\neq 0$. In a similar vein one can define
more general conditional probabilities. For example instead of a
definite value one can allow the respective values to lie in certain
sets, or one can define
\begin{equation}pr(X_i=x_i|\{\mcal{M}_{\nu}\ni Y_j=y_j\})   \end{equation}
with $\mcal{M}_{\nu}$ a subset of boundary random variables.
\end{defi}

We know the following. 
\begin{ob}\label{holo}Denoting by $\mcal{M}_V=\{X_i\}_V$ the full set
  of elementary random variables in $V$, that is, the full set of
  local fluctuations in $V$, and by $\mcal{M}_{\partial
    V}=\{Y_j\}_{\partial V}$ the full set of elementary random
  variables on the boundary, we have
\begin{equation}pr(X_i=x|\{\mcal{M}_{\nu}\ni Y_j=y_j\})\to \delta (x-x_0)
\end{equation}
for a sequence of increasing sets $\{\mcal{M}_{\nu}\ni Y_j=y_j\}$ with
$\mcal{M}_{\nu}\to \mcal{M}_{\partial V}$ and
\begin{equation}x_0=f_i((\{y_j\}_{\partial V})      \end{equation}
\end{ob}
\begin{defi}We abbreviate the above conditional probability distribution
by
\begin{equation}d_{\nu}(x):= pr(X_i=x|\{\mcal{M}_{\nu}\ni Y_j=y_j\})    \end{equation}
(omitting some of the indices for notational convenience)
\end{defi}

Now the following will happen. For each member of the
increasing sets $(\mcal{M}_{\nu},\{y_j\})$ we can calculate the
expected value of $X_i$ and its variance, i.e.
\begin{equation}\overline{x}_{\nu}:=\sum_x x\cdot d_{\nu}(x)\quad ,
  \quad \Delta_{\nu}x^2:=\sum_x (x-\overline{x}_{\nu})^2\cdot  d_{\nu}(x)      \end{equation}
The values $\overline{x}_{\nu}$ will approach the above limit value
$x_0$ while the mean square deviation will become smaller and smaller
with increasing $\nu$ and vanishes in the limit. In other words, the
whole distribution becomes more and more concentrated around
$x_0$.\\[0.3cm]
Remark: Note that due to the huge number of constituents in each
macroscopic volume (compared to the Planck scale) we expect to observe
an almost continuous behavior.\vspace{0.3cm}

As we already remarked above, our aim is to quantify the influence a
DoF on, say, the boundary, $\partial V$, exerts on a given fixed DoF
in the interior, $V$, and, in particular, its dependence on the
spatial distance between the DoF. In this enterprise we only want to
use the holographic information we described above (observation
\ref{holo}) plus very few, as we think, well-motivated simplifying,
assumptions. The natural candidates are the \tit{conditional
  probability distributions}, $d_{\nu}(x)$, which we can (at least in
principle; on our macroscopic scale such observations or measurements
can at the moment presumably not yet be carried out) compare with
$d(x)$, i.e. the unconditioned probability distribution (that is,
without making any assumptions about possible outcomes of observations
of boundary-DoF).  The idea is to study the change of $d_{\nu}(x)$
when the sequence of increasing sets, $(\mcal{M}_{\nu},\{y_j\})$,
approaches the limit set, $(\mcal{M}_{\partial V},\{y_j\}_{\partial
  V})$.

It is reasonable to add (in a thought experiment!) one DoF,
$(Y_j,y_j)$ after another, i.e.
\begin{equation}\label{seq}(Y_1;y_1)\to (Y_1,Y_2;y_1,y_2)\to \ldots \to
  (Y_1,\ldots,Y_{N_{\partial V}});y_1,\ldots,y_{N_{\partial V}}     \end{equation}
Furthermore, we will choose a simple geometric se-tup, i.e. we assume
that $V$ is a ball, $B_R$, of radius $R$, centered at the origin and
hence $\partial V$ a sphere, $S_R$ of radius $R$.

The functions $d_{\nu}(x)$, considered as a whole, have however the
disadvantage that they will move around in the respective variable
space (at least for small $\nu$), so that it may be difficult to use
their changing shape as a quantitative! measure of influence. Note
that there exists a, at first glance, natural distance-measure in
function space, namely the $L^2$-norm:
\begin{equation}\parallel d_{\nu}(x)-d_{\mu}(x)\parallel^2:=\sum_x |d_{\nu}(x)-d_{\mu}(x)|^2     \end{equation}
However, while this norm is widely used in general, it does not really
reflect the particular property that the variance vanishes in the
limit, which we think is actually \tit{the} characteristic property
and provides us with a good measure of influence. So we employ the
following, as we think, characteristic measure of influence.
\begin{ob}We assume that the variance, $\Delta_{\nu} x^2$, is a good
candidate for measuring the influence of the boundary on a bulk-DoF.
\end{ob}
From what we have said above, we start from $\Delta_{\nu=0}
x^2=\Delta x^2$, i.e. the variance with unconstrained boundary-DoF's and
end up with $\Delta_{N_{\partial V}}x^2=0$, that is, with all
boundary-DoF's fixed.\\[0.3cm]
Remark: Note that in our analysis the random variables are assumed to
be discrete, i.e. $N_{\partial V}$ is a very large but finite
number.\vspace{0.3cm}

Now the following seems to be natural while it cannot of course be rigorously
proved. With $N_{\partial V}$ large and
\begin{equation}\Delta_{N_{\partial V}}x^2-\Delta x^2=-\Delta x^2     \end{equation}
a finite number, each DoF, $(Y_j,y_j)$, will add a small amount to the
vanishing of the variance in the limit when we perform the process
described in equation (\ref{seq}). We have however to take the
following into account. Taking in the general case three events
$A,B,C$ the following holds
\begin{equation}pr(A\cap B\cap C)=pr(A|B\cap C)\cdot pr(B|C)\cdot pr(C)     \end{equation}
Only in the case where $B$ and $C$ are statistically independent do we
have 
\begin{equation}pr(B|C)=pr(B)     \end{equation}
Applied to our case we have for example
\begin{multline}pr(X=x,Y_1=y_1,Y_2=y_2)=\\ pr(X=x|Y_1=y_1,Y_2=y_2)\cdot pr(Y_2=y_2|Y_1=y_1)\cdot pr(Y_1=y_1)     \end{multline}
and in general
\begin{equation}pr(Y_2=y_2|Y_1=y_1)\neq pr(Y_2=y_2) \end{equation}
In other words, if we fix $Y_1$ this will also have an effect on the
next boundary-DoF $Y_2$ in our arbitrary but fixed selection, the
value distribution of which will no longer be completely independent
of the previous elements. This means, the summed effect of a large
number of such boundary-DoF on an interior DoF may be somewhat more
involved as in the case of complete independence. On the other hand,
we are only interested in the average influence and, in particular, on
the spatial dependence of this influence. Therefore we proceed in the
following way.

We make the following, as we think reasonable, assumption.
\begin{assumption}We assume that, starting from the original
  unconstrained variance, $\Delta x^2$, of some arbitrary but fixed
  DoF in the interior, $V$, we can represent its ultimate vanishing
  after all boundary-DoF have been fixed by a sum over the individual
  influences or effects of these DoF, modified by a correction term
  which (see above) depends on the position in the selection process
  of the respective boundary-DoF. I.e. we write
\begin{equation}\Delta x^2=\sum_{(Y_j,y_j)} I(|\mbf{r}_j-\mbf{r}_i|;\xi_j)       \end{equation}
The meaning of the various terms is the following: We call $I$ an
\tit{influence function}. It depends on the distance between the
respective DoF on the boundary and the DoF in the interior and on a
set of parameters, abbreviated by $\xi$, which encode its position in
the selection process as indicated above. These parameters contain in
particular the distances of $Y_j$ to the preceeding boundary-DoF in
the selection. We make then the further simplifying assumption that we
are allowed to extract the spatial dependence of
$I(|\mbf{r}_j-\mbf{r}_i|;\xi_j)$ on $|\mbf{r}_j-\mbf{r}_i|$ and write
\begin{equation}I(|\mbf{r}_j-\mbf{r}_i|;\xi_j)=I_1(|\mbf{r}_j-\mbf{r}_i|)\cdot
g_j(\xi_j)     \end{equation}
\end{assumption}
Remark: It is the function $I_1(|\mbf{r}_j-\mbf{r}_i|)$ we are mainly
interested in. It represents the individual influence of $Y_j$ on
$X_i$ when no other boundary-DoF have been fixed. So it comes nearest
to what one views as correlation.\vspace{0.3cm}

Note now that the whole microscopic and presumably messy details of
the behavior of the above sum are contained in the functions
$g_j(\xi_j)$ while the spatial part, $I_1(|\mbf{r}_j-\mbf{r}_i|)$,
should be quite robust. In the $g_j(\xi_j)$ is, among other things,
also encoded the details of the behavior on the chosen values
$\{y_l\}$ of the DoF preceeding $Y_j$. All these details should only
marginally affect the gross spatial behavior. So we feel encouraged to
write
\begin{equation}\sum_{\mbf{r}_j\in
    S_R}I_1(|\mbf{r}_j-\mbf{r}_i|)=O(\Delta x^2)     \end{equation}
 with the number of $\mbf{r}_j$ being proportional to the area of the
 sphere $S_R$. Due to the huge number of involved DoF we can go over
 to an integral and get
\begin{conclusion}The relevant numerical relation is
\begin{equation}\int_{S_R}I_1(|\mbf{r}-\mbf{r}_0|)\,do=O(\Delta x^2)     \end{equation}
with $\mbf{r}_0$ some point in the interior.
\end{conclusion}

We are interested in the decay behavior of $I_1(|\mbf{s}|)$ for large
$|\mbf{s}|$. As usual one makes an ansatz like
\begin{equation}I_1(|\mbf{s}|)=|\mbf{s}|^{-\alpha}\cdot i(\mbf{s})
\end{equation} for non-vanishing $\mbf{s}$ with $i(\mbf{s})$ being
some numerical, bounded and non-decaying function. One could equally
well make e.g. the choice
\begin{equation}|\mbf{s}|^{-\alpha}\to(1+|\mbf{s}|^{\alpha})^{-1}
\end{equation} to avoid a singularity for vanishing $\mbf{s}$, but
such details are not really important. Our central observation is now
the following.
\begin{ob}In
\begin{equation}\int_{S_R}I_1(|\mbf{r}-\mbf{r}_0|)\,do=O(\Delta x^2)     \end{equation}
the rhs is essentially a finite number not depending on $R$ as it
represents the variance of some unconstrained bulk DoF, while the lhs
may in principle depend on $R$. For large $R$ we hence can infer
\begin{equation}\int_{S_R}(|\mbf{r}-\mbf{r}_0|)^{-\alpha}do \approx const    \end{equation}
independent of the radius $R$.
\end{ob}

Due to the inherent symmetry we can make the special choice
$\mbf{r}_0=(0,0,z_0)$ with $z_0=R\cdot k$ and $0\leq k\leq k_0<1$.
We have to evaluate the integral (in three space dimensions):
\begin{equation}R^2\cdot\int_0^{2\pi}d\phi\int_{-1}^{+1}d\!\cos\!\theta\,
  (x^2+y^2+(z-z_0)^2)^{-\alpha/2}  \end{equation}
($x=R\sin\theta\sin\phi,\ldots$)
With the above choice for $z_0$ this yields
\begin{equation}2\pi\cdot R^{(2-\alpha)}\cdot\int_{-1}^{+1}du\,((1+k^2)-2ku)^{-\alpha/2}      \end{equation}
Remark: The integrand is always positive. This follows already from
the structure of the integrand, we started from. On the other hand, we
have
\begin{equation}(1+k^2)-2ku=(1-k)^2+(2k-2ku)>0      \end{equation}
as $k^2<1$ and $2k-2ku\geq 0$ since $|u|\leq 1$.
\begin{conclusion}We see that we can only avoid a contradiction if we
  have in leading order
\begin{equation}\alpha=2     \end{equation}
This implies
\begin{equation}I_1(|\mbf{r}-\mbf{r}_0|)\approx  |\mbf{r}-\mbf{r}_0|^{-2}   \end{equation}
in three space dimensions and analogous results in other
dimensions. In other words, the influence decays like a Coulomb
force-law for sufficiently large distances.
\end{conclusion}

For later use we can calculate the integrated influence of a sphere on
a DoF in the exterior, i.e. for $k\geq k_0>1$. For $\alpha=2$ the integral
\begin{equation}\int_{-1}^{+1}du((1+k^2)-2ku)^{-1}     \end{equation}
is dominated for large $k$ by the first term in the denominator. We
thus have
\begin{koro}For points outside the sphere $S_R$ with (for convenience
  $\mbf{r}_0=(0,0,k\cdot R)$) the above integral decays as $\sim
  k^{-2}$ for large distances $k$ from the sphere.
\end{koro}
This has important consequences which will be discussed in section
\ref{appli}.

It is perhaps worth mentioning that a clustering of correlations
$\sim|\mbf{x}|^{-2}$ in four space-time dimensions occurs also in the
context of vacuum-Bell-inequalities in \cite{Werner}, theorem 4.1 or
in massless quantum field theories (cf. \cite{Araki}).

\subsection{\label{4.2}Long-Range Correlations and Goldstone Phenomenon in
  Ordinary Physics}
In this subsection we want to show by means of counterexamples that
lon-range (anti)correlations alone are not sufficient to entail an
area-law behavior of entropy. It turns out that frequently the
underlying reason seems to be the existence of Goldstone modes or
other collective excitations, representing small fluctuations around
the ordered state.

In \cite{Requ1} we discussed in some detail the example of the
(3-dimensional) harmonic crystal in a \tit{pure phase}. This means in
this context that its global position is assumed to be fixed in space
and does not fluctuate. In this and related systems we observe the
phenomenon of spontaneous symmetry breaking of a continuous group. In
this particular case the translation group is broken with the
\tit{phonons} as \tit{Goldstone-modes}. In such a situation we have
long-range (anti)correlations between certain observables, in our case
these are the atomic positions and the respective deviations from
their equilibrium positions .

The crucial estimate in \cite{Requ1} having a certain bearing on our
present discussion is equation (35) in section 4, the meaning of which
is the following. With the global position of the crystal being fixed,
the fluctuations of the individual atomic positions can be bounded by
\begin{equation}\delta x_i^2:= \langle (x_i-<x_i>)^2\rangle\lesssim
  a^2\label{a}\end{equation}
with $a$ the lattice spacing. For simplicity we take a row of $j$ atoms
on, say, the x-axis, starting from $x_0$ and ending in $x_j$. Denoting
$(x_k-x_{k-1})$ by $u_k$ with $<u_k>=a$, we showed in \cite{Requ1}
that due to (\ref{a}) we can infer that
\begin{equation}\langle\sum_{k\neq
    k'=1}^j(u_k-a)(u_{k'}-a)\rangle\approx -
  \langle\sum_{k=1}^j(u_k-a)^2)\rangle\lesssim-j\cdot (2a)^2
\end{equation}

In other words, the left hand side of the equation which contains a
double sum over the correlations of the relative atomic positions,
$u_k$, is strongly negative, meaning that a positive elongation at
site $k$ has to be compensated by negative elongations at other sites
so that the fluctuation of $x_j$ which is essentially a sum over all
the $u_k$ remains small. This is an example of long-range
anticorrelation.

Nevertheless, irrespective of the presence of this long-range order
the entropy of subvolumes is proportional to the volume and not! the
area of the boundary. That is, while it is clear that an ordered phase
has an entropy which is lower than in the unordered phase, it is still
an extensive quantity.\\[0.3cm]
Remark: Note that the standard thermodynamic explanation for the
(almost) additivity of entropy relies on short-range correlations and
the possibility to divide a large system into a (large) number of
weakly interacting small but still macroscopic subsystems. In the
presence of long-range correlations this is no longer
possible.\vspace{0.3cm}

The (perhaps surprising) deeper reason for this volume behavior of
e.g. the entropy in the face of long-range order can be understood in
our view in the following way. Phase transitions as a consequence of
some spontaneous symmetry breaking fall more or less in two broad
classes.  First, the class with the broken symmetry belonging to a
continuous (Lie-) group. Second, the class where the symmetry happens
to be discrete, a typical example being the Ising model. In the first
case, the \tit{Goldstone theorem} (in fact the consequence of the
spontaneous breaking of a continuous symmetry) tells us that, at least
in the case of short-range interactions, there exist \tit{gapless}
Goldstone modes, i.e. modes whith the energy-momentum dispersion law,
$\varepsilon(\mbf{p})$, passing
through $\varepsilon=0$ for $\mbf{p}=0$.\\[0.3cm]
Remark: Strictly speaking, in the non-relativistic regime these modes
usually have, except in the interaction-free case, a finite lifetime,
i.e. a certain $\mbf{p}$-dependent width. We neglect these details in
the following.\vspace{0.3cm}

Employing the Landau picture of elementary excitations in many-body
theory (see e.g. \cite{Abri} or \cite{Mattuck}), meaning that
approximately the Hamiltonian or the excitation spectrum can be
understood as a sum over weakly interacting collective or elementary
excitations ( the ``normal modes'' of the system), we can write
\begin{equation}H\approx \sum_i H_i\quad\text{with}\quad H_i:=\int
  \varepsilon_i(\mbf{p})\cdot a_i^{\dagger}(\mbf{p})a_i(\mbf{p})\,d^n\!p  \end{equation}
We can now infer that the total entropy, $S$, is roughly
\begin{equation}S\approx \sum_i S_i\quad\text{or at least}\quad S\geq
  \sum_i S_i \end{equation}
 with $S_i$, belonging to the free system
built over $\varepsilon_i(\mbf{p})$, being extensive, i.e. being
proportional to the volume. In our case there does exist at least one
such excitation branch (the Goldstone excitations). This means that
while there is of course an increase of order in the system below a
phase transition point or line, resulting in a decrease of entropy,
the latter is still homogeneous in the volume as it contains the
contributions of basically free systems.

In our above example the Goldstone modes are the phonons, the
approximative system is a free gas of phonons. In classical terms,
they represent the normal coordinates belonging to collective
oscillation modes of the lattice atoms about their equilibrium
positions. To make a clearer connection to our BH-topic, we may assume
that for example the atoms on the boundary of a macroscopic subvolume
are held fixed. But nevertheless, due to the fact that there can
exist phonons with arbitrarily low energy (small oscillations), for
non-vanishing temperature $T$, the modes belonging to the enclosed
bulk system fill a full region of phase space. That is, the phase
space volume of the interior bulk system with the boundary values
held fixed still fulfills
\begin{equation}\Omega_V^{bd}\sim e^V\quad\text{i.e.}\quad S\sim V    \end{equation} 

This latter procedure is for example one of the methods to generate a
\tit{pure phase} in the bulk interior. In the case of a spin system
like e.g..
\begin{equation}H(S)=\sum_{ik}J_{ik}\mbf{S}_i\cdot\mbf{S}_j
\end{equation}
one procedes by fixing the spins on the boundary of some large
subvolume and study the interior system in, say, the limit
$V\to\infty$. Below a critical point the system will display some
preferred direction of magnetisation (or some other kind of order
parameter), that is, develop \tit{long-range order} with the
thermodynamic entropy being proportionalto the volume. The gapless
excitations are the \tit{magnons} in this case.

That our explanation is reasonable, can be tested by analysing the
situation for the second class of systems
displaying spontaneous symmetry breaking. A typical representative is
the Ising model with the spins having for example the two orientations
$\pm 1$. The broken symmetry is now \tit{discrete}; consequently there
are in general no gapless Goldstone excitations. On the other hand,
one can show (\cite{He},\cite{Fisher},\cite{Lebo}) that below the
critical point (the realm of spontaneous magnetisation) the truncated
correlation functions decay exponentially. That is, we have the
situation, discussed earlier, of short-range correlated systems with
the usual volume behavior of entropy.

There are a few examples where systems develop a gap due to long-range
interactions (e.g. Coulomb). Another case in point is the famous
BCS-model of superconductivity. In this example we have long-range
correlations and Goldstone modes not passing through $E=0$ for
$\mbf{p}=0$. However, in all of these cases the energy gap above the
groundstate is of atomic size and in general much smaller than the
energy scale given by the temperature $T$. As in the canonical or
grandcanonical ensemble, the energy is allowed to stretch over, in
principle, arbitrary scales, entropy is still linear in the volume.
Matters may become different however, if these two scales become
comparable. This last observation leads over with almost necessity to
the topic adressed in the next subsection.
\subsection{\label{4.3}Systems with a large Energy Gap}
In subsection 4.1 we analysed the fine structure of the long-range
anticorrelations which show up in connection with the holographic
principle in our approach. Now we provide a deeper physical reason for
this strange and counterintuitive behavior in form of a dynamical or
spectral property of certain Hamiltonians which may play a role in
this field. 

We have finally (cf. subsection 4.2) located the essential ingredient
which was still missing.
\begin{conjecture}The crucial precondition for systems of statistical
  mechanics having an entropy which is proportional to the area of the
  bounding surface is a sufficiently large gap in the energy spectrum
  of the bulk-Hamiltonian, defined on $V$ with fixed boundary
  conditions on $\partial V$, above the ground state energy or between
  a few low-lying excited states and the rest of the energy-spectrum.
  More precisely, the gap has to be so large that it exceeds the
  typical excitation energies being considered to be present in the
  respective situation.
\end{conjecture}
That is, given a system-Hamiltonian, $H_V$, over the sub-volume $V$
with, for reasons of simplicity, discrete spectrum,
$\{\varepsilon_i\}$, we assume that
\begin{equation}\varepsilon_0\leq\varepsilon_1\leq\ldots
  \leq\varepsilon_N\ll
  \varepsilon_{(N+1)}\leq\varepsilon_{(N+2)}\ldots  \end{equation}
I.e., we assume a large energy gap,
$\Delta:=\varepsilon_{(N+1)}-\varepsilon_N$ between the energy values
$\varepsilon_N$ and $\varepsilon_{(N+1)}$ with $N=O(1)$, which is so
large that the typical energies, $E$, existing in the system fulfill
\begin{equation}E<\varepsilon_{(N+1)} \end{equation} Physically this
means that in the usual situation only finitely many energy levels are
occupied while in principle the total number of eigenvalues of $H_V$
may nevertheless be infinite.

This observation needs however some more specifications. A Hamiltonian
defined over a finite volume, $V$, needs for its complete
specification \tit{boundary conditions} on its bounding surface,
$\partial V$. That is, we state the following:
\begin{ob}For our purpose Hamiltonians, $H_V^{bd}$, are relevant which for
  each selected boundary condition, $bd$, on $\partial V$ out of a
  class of admissible conditions, $Bd$, have spectral properties as
  assumed above, i.e.  
\begin{equation}\varepsilon_0^{bd}\leq\varepsilon_1^{bd}\leq\ldots
  \leq\varepsilon_{N_{bd}}^{bd}\ll
  \varepsilon_{(N+1)_{bd}}^{bd}\leq\varepsilon_{(N+2)_{bd}}^{bd}\ldots  \end{equation}
with $N_{bd}=O(1)$ and
$\Delta_{H_V^{bd}}=\varepsilon_{(N+1)_{bd}}^{bd}-\varepsilon_{N_{bd}}^{bd}>E$
uniformly in $bd\in Bd$, with $E$ some typical energy scale available
in the respective scenario.
\end{ob}
The question is of course, do there exist physical mechanisms which
generate such a peculiar spectral behavior?

It turns out that a closer inspection of this problem leads to
surprising consequences and ramifications which go far beyond the, at
first glance, seemingly isolated technical question we are posing. To
put it briefly, we found that behind this question is lurking the
question of the microscopic (causal) organisation of our space-time.
Or stated differently, what is called for are new types of \tit{space
  forms} which go substantially beyond our ordinary classical
(continuum) geometries.

We should note that the existence or necessity of energy gaps shows up
also elsewhere in modern high-energy physics. In \tit{Kaluza-Klein
  theories}, for example, we have a tower of widely separated energy
scales, labelled by the excitation modes of the small internal space.
In supersymmetry it is also argued that we do not see (most of) the
supersymmetric partners because they are so heavy.

But here the situation is different and much more involved. In our
case the distribution of energy (eigen)values in the bulk follows a
surface-law because the effect of the surface is strongly felt in the
interior. Put differently, the underlying reason is not some internal
small space but rather the \tit{entangled} or in some sense
\tit{non-local} microscopic fine structure of the real geometric
space-time.

The reason is the following. It is obvious that the holographic
principle introduces a specific kind of \tit{quantum non-locality}
into the framework which seems to extend the many forms of
``non-locality'', which are almost ubiquituous in ordinary quantum
theory (whereas they are frequently ``discussed away'' as it is felt
that they are in conflict with the locality dogma). The most prominent
is the pure quantum phenomenon of \tit{entanglement}.
\begin{conjecture}We conjecture that the kind of quantum
  non-locality, observed in the area-law of BH-physics and the
  holographic principle, and the various aspects of non-locality and
  (long-range) entanglement being present in quantum theory as such,
  are of exactly the same nature. They are both the result of a
  particular non-local (with respect to the classical realm of
  space-time!) microscopic organisation of quantum space-time.
\end{conjecture}
We described the deep structure of space-time in recent work in more
detail (see e.g. \cite{Requ3},\cite{Requ-Gromov},\cite{Requ-Wormhole}
and further references given there). One should also note that closely
related phenomena do occur in the \tit{small-world scenario} (see e.g.
\cite{Requ-Small} for a rigorous discussion and more references).

Recently we came upon a promising geometric generalisation of
Riemannian geometry in pure mathematics, called \tit{subriemannian
  geometry} or \tit{Carnot-Caratheodory spaces}, which seems to be
able to encode some aspects of this geometric ``double structure'' we
have in mind and which we called \tit{wormhole spaces}. That is, a
classical continuum surface structure with an ordinary distance metric
embedded in an ambient \tit{hyperspace} which allows for
\tit{short-cuts} between classically widely separated regions (cf.
e.g. \cite{subriemannian1} or \cite{subriemannian2}).

On the other hand, we have to provide arguments that the long-range
entanglement structure, encoded in the microscopic structure of
space-time (at or near the Planck level) has as one of its
consequences this mentioned \tit{gap-structure} in the excitation
spectrum. As this requires a quite extensive (and technical)
investigation of its own we will give the necessary details elsewhere. 
\section{\label{interior}The Interior of the BH on the Microscopic Scale}
From the results of the preceding section we can now develop a picture
of the microscopic state of a BH (for reasons of simplicity we only
discuss the Schwartzschild-BH). We remarked several times in this
paper that we regard our framework among other things as an extension
and generalisation of old ideas of Sakharov, Zeldovich et al
(\tit{induced gravity}). The BH-interior is in our view a particularly
instructive example.

In contrast to ordinary regions of, for example, the Minkowski vacuum,
the characteristic property of the BH-interior on a microscopic scale
is in our approach that the vacuum fluctuation structure is deformed
by the central singularity. While macroscopically this singularity is
essentially structureless (the only characteristic being the central
mass, $M$), we think that the microscopic deformation structure of the
vacuum fluctuation pattern in the BH-interior expresses the history of
the formation of the BH. This was already mentioned as a possible
source of BH-entropy in the literature. Our microscopic analysis
supports this point of view.

We will come to the interesting possibility of relating the
macroscopic (classical) solutions of the Einstein equations to the
corresponding microscopic vacuum fluctuation patterns in subsection
\ref{range}. As to the BH-solution we can formulate the following
conclusion
\begin{conclusion}From our analysis the following picture does
  naturally emerge. The different possible ways of creating the same
  macroscopic BH are expressed in the different microscopic
  fluctuation patterns existing in the BH-interior. An observer being
  located in the exterior of the static BH has in general no knowledge of
  this microscopic fine structure. It follows that for him all these
  microstates have the same probability. This implies that for him the
  BH is in a state of maximal entropy.
\end{conclusion}
\section{\label{mesoscopic}Mesoscopic Excitation Patterns and their Correlations}
In the introductory sections we argued that the fluctuation spectrum
has to be long-range (anti)correlated. Otherwise it would be possible
to observe the integrated fluctuations in macroscopic subvolumes,
which does not conform with our general experience. One should note
that this argument applies, in the first place, to the state we call
vacuum on a macroscopic or mesoscopic scale; put differently, to the
ensemble of microscopic excitation patterns which look macroscopically
like the vacuum. Here we employ the same philosophy as in statistical
mechanics, where whole subclasses of microstates are subsumed under
the corresponding macrostates.

The situation is slightly different for states containing particle
excitations. These particle excitation patterns are extended regions of small
but coherent deformations of the vacuum fluctuation structure and last
for macroscopic times. Here we share the working philosophy of
e.g. Sakharov mentioned above. But according to the general
\tit{holographic philosophy} and due to the long-range correlations we
descibed in the preceding sections, every particle excitation,
localized for example deep in the interior of $V$, has its counterpart
in form of a characteristic excitation pattern on the boundary of $V$.   
\begin{ob}From the above argument of a correspondence of an internal
  particle excitation and a unique boundary excitation (microscopic
  one-one correspondence or with respect to equivalence classes) it
  follows that only a certain amount of particles can be stored in a
  finite volume, with this number depending on the area of the
  bounding surface. More specifically, an $N$-particle state,
  mesoscopically localized in the interior, has to be expressed by a
  boundary excitation pattern which corresponds uniquely to this
  interior state (or in the sense of classes of microstates). But the
  number of different possible boundary states is proportional to the
  area of the boundary.
\end{ob}

In connection with such mesoscopic (or macroscopic) excitation
patterns there exists another interesting question which has to be
clarified in order to show the consistency of the framework. We
learned that on the primordial level the respective fluctuations are
long-range correlated. We argued that the spatial influence decays
like $\sim |\mbf{r}|^{-2}$ in three space dimensions. On the other
hand, in mesoscopic or macroscopic model theories, with constituents
such particle-like excitations, we observe on these scales of lesser
resolution of space-time all possible kinds of decay of correlations,
from short-range to long-range. That is, we have for instance to
explain how on a coarser scale short-range correlations between the
respective constituents of a model theory do emerge from long-range
correlation among the elementary DoF on a microscopic scale.
\begin{ob}The task consists of explaining a partial decoupling between
  these different scales of resolution. Note that compared to e.g. the
  Planck scale all ordinary scales are of macroscopic size. As may be
  expected, the decoupling will be a result of a certain
  coarse-graining and averaging-out of finer details.
\end{ob}

To discuss this problem in a more concise form we introduce some notation.
\begin{defi}The subset of microstates, being compatible with the
  macroscopic state we call vacuum we denote by $\Omega_{vac}$
\begin{equation}\Omega\supset \Omega_{vac}=\{\omega_j^{vac}\}
\end{equation}
with $\Omega$ the total set of possible microstates. By the same token
the set of microstates belonging to a certain particle state, $\psi$
(e.g. a wave function), is denoted by 
\begin{equation}\Omega_{\psi}=\{\omega_j^{\psi}\}      \end{equation}
etc. Alternatively we denote these subsets or classes by
$[vac]\;,\;[\psi]$ etc.
\end{defi}

It is easier to discuss the notion of correlations in the realm of
statistical mechanics and ensembles. That means, we have a macrostate
consisting of a number, $N$, of particles confined to some volume in
space (e.g. a temperature state). To discuss correlations one
typically uses the ensemble picture. In other words we assume to have
an ensemble, $\Lambda=\{\psi_i^N\}$, of $N$-particle (quantum)-states
belonging to the macroscopic state. Now, on the two levels of
resolution, the mesoscopic or macroscopic level, denoted by $II$, and
the microscopic level, denoted by $I$, the situation is described in
the following way.

Let us for example assume that we want to describe the correlations
between the individual localisations (positions) of the $N$ particles.
This is a natural concept in statistical mechanics. Each $\psi_i^N$ on
level $II$ comprises a class of microscopic configurations on level
$I$, i.e.
\begin{equation}[\psi_i^N]=\{\omega_j^{\psi_i}\}      \end{equation}
On level $I$ the mesoscopic ensemble, $\Lambda$, is represented by
\begin{equation}\Lambda=\{\psi_i^N\}=\{[\psi_i^N]\}       \end{equation}
Remark: We can as well discuss correlations in pure quantum
many-particles states. Then we start directly from states like
$\psi^N$ instead of ensembles like $\Lambda$.\vspace{0.3cm}

The important point is that in the model theories on level $II$ the
microscopic fine structure, being present in the primordial states,
$\omega_j$, is not encoded or at most in an averaged sense. This holds
in particular for the long-range nature of the correlations on level
$I$ and the correspondence of bulk and boundary excitations. One can
state this in a possibly more illuminating way. The observables on
level $II$ and level $I$ are entirely different and only loosely
coupled to each other. While on level $II$ we are for example
interested in positions of particles, i.e. weak large-scale
deformations of fluctuation patterns on level $I$, and their
correlations, the long-range correlations on level $I$ exist mainly
between the individual elementary fluctuations or DoF. These latter
correlations are averaged or washed out in the transition from level
$I$ to level $II$.
\begin{conclusion} From the above it follows that the correlation
  structures of level $I$ and level $II$ are decoupled to a large
  extent or are related only in a relativley subtle way. As to this
  latter point we think e.g. of the phenomenon of entanglement and
  similar quantum phenomena.
\end{conclusion}  
\section{\label{appli}Applications}
In this section we want to apply our framework to a variety of issues
and problems which have been raised in connection with the holographic
principle and related entropy bounds, with the aim of supplying
answers from our point of view or illuminate the problem under
discussion from another angle. The topics to be discussed will be: i)
the existence of a \tit{natural} cutoff, ii) the species problem, iii)
cosmological backgrounds like closed-static or time-dependent which
seem to imply that the simple spatial holographic principle has to be
generalized, iv) the problem of unitarity.
\subsection{The Problem of a Natural Cutoff}
This problem becomes particularly virulent in practically all
approaches which use quantum field theoretic methods. We do not intend
to give a review of this problem as it is more or less ubiquitous (see
for example \cite{Hooft2}, a catchword being \tit{brickwall}). Usually
such a high energy cutoff is introduced in a relatively adhoc manner
whereas frequently the Planck scale is invoked in this context. It is
necessary to render many continuum calculations finite, in particular
the entropy itself.

We surmise however that these continuum-field calculations are only
approximations, being correct only for low energies (i.e. exactly the
opposite end of the usual scenario). The findings of our present paper
are perhaps able to make the existence of such a cutoff more natural.
We remind the reader that our starting point was the observation,
almost rigorously proved (at least by physical standards), that by
necessity the fluctuation spectrum in the (quantum) vacuum (of energy,
momentum or what else) is both long-range and strongly anticorrelated
on microscopic scales (which tacitly means, according to general
folklore, the Planck scale).

That means that on already very small scales positive and negative
deviations from some average, representing the macroscopic vacuum,
have to compensate each other (anticorrelation) and that this
compensation pattern is quite rigid over large distances (long-range
correlated). 
\begin{conclusion}In the light of these two characteristics of the
  vacuum fluctuation pattern, it appears to be reasonable to associate
  the typical size of the grains of synchronous fluctuation with the
  Planck scale, thus leading to a relatively natural cutoff in length,
  energy etc.   
\end{conclusion}
\subsection{The Species Problem}
To discuss this point, we can use the notations introduced in section
\ref{mesoscopic}. The species problem is for example addressed in
\cite{Bekenstein2}, section 2.3 or in \cite{Bousso}, section II.C.4,
to mention just a few sources. It consists roughly in the following.
In the standard calculations of the various entropy bounds, made in
the literature, the number of different species of particles play a
role, which are assumed to be confined to some ball, $V$ of radius
$R$. One has for example for a gas of photons the sequence of relations
\begin{equation}R\geq 2E\;,\;E\sim ZR^3T^4\;,\;S\sim ZR^3T^3
                         \end{equation}
with $Z$ the number of particle species, $E$ the thermodynamic energy,
$T$ the temperature and $S$ the entropy. From this one can infer
\begin{equation}S\lesssim Z^{1/4}A^{3/4} \end{equation} (see
\cite{Bousso},loc.cit.)so that we can violate the spherical entropy
bound if $Z$ is greater than $A$. In short, the calculation of material
entropy depends on the number of different particle species, while the
spherical entropy bound is purely geometric.

In our framework this problem is resolved in the following way. With
the basis of everything being the microscopic (Planck-size)
fluctuation pattern of the elementary DoF, we argued in e.g. section
\ref{mesoscopic} that particle/field excitations happen to be
large-scale deviations of this ground pattern. This implies that on a
truely small scale the alleged different particle excitations in such
an extremely densely packed ensemble of particles loose their
individuality which they have on small energy scales and end up in the
general background of excitations of the elementary DoF. That is, on
the microscopic scale there do no longer exist different species.
This then explains the purely geometric character of the area law.
\subsection{\label{range}The Range of Validity of the Spatial Holographic Bound}
In \cite{Bousso} and elsewhere (e.g. \cite{Easther} or
\cite{Fischler}) counter-examples are given, which show that the
simple \tit{spatial holographic bound} in its original purely
geometric form has only a limited range of application and has hence
to be given a more general meaning (\tit{covariant holographic
  principle}). We refer also to the recent \cite{Bru} concerning
cosmological entropy bounds. In the following we want to show that our
approach sheds some new light on the unifying principles underlying
these seemingly different bounds and, a fortiori, allows to draw
important conclusions about the microscopic dynamics, going on in the
deep-structure of the quantum vacuum and its relation to the
cosmological solutions of the Einstein equations. Therefore, our
microscopic analysis may be able to complement the various geometric
extensions of the original, perhaps too narrow, idea.
\subsubsection{The Closed Static Universe}
One counter-example is a space-time, $\mcal{M}$, containing a closed
space-like hypersurface $\mcal{O}$ (see \cite{Bousso}, section
IV.B.1). We divide  $\mcal{O}$ into
\begin{equation}\mcal{O}=V\cup Q          \end{equation}
with $Q$ a small ball-like set in $\mcal{O}$ (and thus having a small
boundary). In $V$ we place a certain macroscopic amount of matter
having a macroscopic entropy. By making $Q$ and thus the common
boundary sufficiently small, the geometric area-law can easily be
violated for the large volume $V$.

Our detailed analysis in section \ref{Influence} shows that the
area-law should not be regarded as a God-given geometric law, coming
somehow from outside, but, quite to the contrary, is the result of a
very subtle microscopic correlation and influence structure between,
for example, bulk and boundary of a region. What is really crucial is
the \tit{influence formula}
\begin{equation}I(|\mbf{r}-\mbf{r}'|)\approx |\mbf{r}-\mbf{r}'|^{-2}
\end{equation}
 in three space dimensions. For a ball like $Q$ with
spherical boundary $S_R$, the area-law holds for the interior of $Q$.
We showed at the end of section \ref{Influence} that and why the
situation is different for points lying outside of $S_R$. 

For $S_r$ concentric with $S_R$ but $r>R$, the DoF on $S_R$ are not
entirely fixed by the configuration on $S_R$ but only happen to be
restricted statistically in their variance and this statistical
influence becomes weaker and weaker with increasing $r$. More
specifically
\begin{conclusion}With $Q$ a small ball in the closed space $\mcal{O}$
  of radius $R$, the area-law in its simple form does not hold in the
  exterior of $Q$, i.e. in $V$. With $B_r$ a ball concentric with $Q$
  and $r\gg R$ we have for the maximum entropies
\begin{equation}S_{max}(B_r\backslash Q)\approx  S_{max}(B_r) -
  S_{max}(Q)\approx S_{max}(B_r)     \end{equation}
 and hence
\begin{equation}S_{max}(V)\geq  S_{max}(B_r\backslash Q)\gg S_{max}(Q)   \end{equation}
\end{conclusion}
That is, in our framework this is physically a rather natural result.
\subsubsection{An Expanding Non-Closed Universe}
Considering our universe as expanding, practically flat and spatially
unbound, we can assume that it is more or less homogeneously filled
with matter of a, on average, low but non-vanishing density and hence,
by the same token, non-vanishing entropy-density. That is,
we can find sufficiently large spheres with an entropy content which
exceeds the value, given by the spatial entropy bound.

In our approach this example and other time-dependent scenarios are
particularly illuminating from a microscopic point of view, underlying
the holographic principle. One should note that in the preceding
section we have mostly dealt with the spatial holographic principle in
approximately static space-times. In such scenarios we derived the
above mentioned microscopic correlation result 
\begin{equation}I(|\mbf{r}-\mbf{r}'|)\approx |\mbf{r}-\mbf{r}'|^{-2}
\end{equation}
On the other hand, we remarked in e.g. section 2 that we regard our
paper (among other things) as an extension and generalisation of old
ideas of Sakharov, Zeldovich and Wheeler (\tit{induced gravity}). We
remind the reader that in those contributions (macroscopic) gravity
was viewed as a derived and secondary effect arising from (or, rather,
being equivalent to) variations and deformations in the vacuum
fluctuation spectrum.

The above observations, being made in the context of time-dependent
cosmological scenarios, now clearly indicate the following. 
\begin{ob}The fluctuation pattern, its correlation structure and
  dynamics in the deep structure of the physical vacuum or space-time
  (that is, on level $I$ (cf. section \ref{mesoscopic})) is standing in
  correspondence to the respective solutions of the Einstein equations
  (on level $II$). Or phrased slightly differently, the classical
  solutions are the coarse-grained picture of their microscopic
  counterparts. If the former are dynamic and time-dependent, the same
  holds for the corresponding  microscopic fluctuation and correlation
  patterns. 
\end{ob}
\begin{koro}In such expanding solutions the correlations on level $I$
  decay, for example, faster than $\sim|\mbf{r}|^{-2}$, thus leading to
  an increased storage capacity of information per volume.
\end{koro} 
\subsection{Unitarity}
In connection with the formation and evaporation of a BH the problem
of unitarity is strongly felt. This problem has been extensively
discussed in the literature. In order not to blow up the
representation too much, we only mention the discussions in
e.g. \cite{Wald1}, p.29ff, \cite{Bousso}, sections III.F,G and
\cite{Susskind_Book}, chapt. 9. The question is basically whether the
information which has gone into the BH (starting for example from some
pure initial state) is still somehow encoded in the thermal spectrum
which is emitted by the BH during its evaporation.

Conventional wisdom tells us that in thermal states information is
lost. However, physical intuition, trained on relatively simple
examples may be deceptive. 
\begin{ob}In general there is no clear apriori distinction between
  pure vector states and mixtures, i.e. density matrices in quantum
  theory.
\end{ob}
This means the following. A unique division between vector states and
density matrices can be made in Hilbert space representations where
the full set of bounded operators, $\mcal{B}(\mcal{H})$, corresponds
to the algebra of observables. In the mathematical classification of
operator algebras (originally given by Murray and v.Neumann) this is
called the type-I case. 

On the other hand, in cases where this is not so, existence of event
horizons, only incomplete local information access etc., there exist
general mathematical statements to the effect that it can happen that
in the same Hilbert space representation each density matrix can be
represented by a pure Hilbert space vector (see e.g. \cite{Haag} or
\cite{Wald2}) for a more physically motivated discussion or
\cite{Sakai}, Lemma 2.7.8, p.104 or \cite{Takesaki}, Corollary 1.12,
p.295, for complete mathematical proofs. Helpful is perhaps also the
discussion in\cite{Sciama}.

In a similar directions points the observation we mentioned in
\cite{Requ2}, section 3. That is, a Gibbs state, given on a subsystem,
$\mcal{H}_1$, can be extended to a pure vector state on an extended
system, $\mcal{H}_1\otimes\mcal{H}_2$ for some suitable $\mcal{H}_2$
with
\begin{equation}\Psi:=\sum \sqrt{p_i}\cdot \psi_i\otimes e_i
\end{equation}
$\{e_i\}$ spanning a basis in $\mcal{H}_2$, $\psi_i$ the eigenstates
of the Hamiltonian in $\mcal{H}_1$ and $p_i$ the Boltzmann weights.

Concerning our particular case of formation and evaporation of a BH,
the question is of course the scale of resolution of space-time we
want to employ and the amount of information a real or hypothetical
observer is granted. For the truely microscopic scale, which interests
us here, we refer to our remarks in section \ref{interior}.

\end{document}